\documentclass{aa}  

\usepackage{graphicx}
\usepackage{txfonts}
\usepackage{color} %%ADDED

\begin{document} 

%%forAstridPeter
 \title{Spatial variations of the Sr~{\sc i} 4607~\AA\ scattering polarization peak}

%   \subtitle{..}

   \author{M. Bianda
          \inst{1}
          \and
          S. Berdyugina
          \inst{2}
          \and
           D. Gisler
          \inst{1,2}
          \and
          R. Ramelli
          \inst{1}
          \and
          L. Belluzzi
          \inst{1,2}
          \and
          E.S. Carlin
          \inst{1}
         \and
         J.O. Stenflo
          \inst{1,3}
          \and
         T. Berkefeld \inst{2}
          }

   \institute{Istituto Ricerche Solari Locarno, IRSOL, Locarno, associated to
     Universit\`a della Svizzera Italiana, Switzerland,
      \email{mbianda@irsol.ch}
         \and
         Kiepenheuer Institut f\"ur Sonnenforschung, Freiburg, Germany   
         \and
         ETHZ, Zurich, Switzerland
}

%  \date{Received September 15, 1996; accepted March 16, 1997}

% \abstract{}{}{}{}{} 
% 5 {} token are mandatory
 
  \abstract
{
%\LEt{this is not mandatory, but I suggest a slight rephrasing
%of your title to avoid piling on the adjectives so: Spatial variations
%in the scattering polarization peak of Sr I 4607 A}
The scattering polarization signal observed in the photospheric Sr~{\sc i}
4607~\AA\ line  is expected to vary at granular spatial scales. 
This variation can be due to changes in the magnetic field intensity and 
orientation (Hanle effect), but also to spatial and temporal variations in
the plasma properties. 
Measuring the spatial variation of such polarization signal
would allow us to study the properties of the magnetic fields at subgranular 
scales, but observations are challenging since both high spatial resolution 
and high spectropolarimetric sensitivity are required.  
}
{ 
We aim to provide observational evidence of the 
polarization peak spatial variations,  
and to analyze the correlation they might have with granulation. 
}
{
Observations conjugating high spatial resolution and high 
spectropolarimetric precision were performed with 
the Zurich IMaging POLarimeter, ZIMPOL, at the GREGOR
solar telescope, taking advantage of the
adaptive optics system and the newly installed image derotator.
}
{
Spatial variations 
of 
the scattering polarization in the Sr~{\sc i}
4607~\AA\ line are clearly observed.
The spatial scale of these variations is comparable with the granular size.
Small correlations  between the polarization signal amplitude and the 
continuum intensity 
indicate that the polarization is higher at the center of granules than in
  the intergranular lanes.
 }
   {}


   \maketitle

%-------------------------------------------------------------------

\section{Introduction}

The Sr~{\sc i}  4607~\AA\ line shows one of the strongest scattering
polarization  signals
in the visible solar spectrum; this line was observed for the first time 
at Istituto Ricerche Solari Locarno, IRSOL, 
by \citet{wiehr-1975}.
On the basis of 3D simulations, \citet{trujillobueno-2007} predicted spatial 
variations in the amplitude of this signal, provided
that the granulation were resolved. 
These fluctuations originate from local variations in the anisotropy  
of the radiation field,
 as well as from variations in the magnetic field
between granule interiors and intergranular lanes.
Several attempts to observe this behavior have been made, and a first
indication of  differential depolarization in granular interiors and
intergranular lanes was  found by \citet{malherbe-2007}. 
Using the Zurich IMaging POLarimeter, ZIMPOL,
\citep{ramelli-2010} at the Gregory Coud\'e telescope at IRSOL
(45~cm aperture), we were only able to measure sporadic spatial variations of
the linear polarization peak of the Sr~{\sc i} 4607~\AA\ line, at scales well
above the granular  one.
Other observations are currently performed  with the same aim by a group at
the Max Planck Institute for Solar System Research; this group has 
obtained positive results (F. Zeuner and A. Feller, private communication). 

We report here the results of the observing campaign we performed in 2016 at
the GREGOR
solar telescope in Tenerife \citep{Schmidt_2012}, using the ZIMPOL
polarimeter \citep{ramelli-2014}.
The seeing conditions during the campaign allowed us to achieve the required
spatial resolution  during only a few hours, when we chose to observe a quiet
region near the East limb.
We could reach a subarcsecond spatial resolution
(we estimate about $0.6\arcsec$), which 
allowed us to detect the expected fluctuations in the linear polarization.

Hereafter we define as positive Stokes $Q$ the linear polarization 
parallel to the closest solar limb.
In addition to spatial variations in the amplitude of the $Q/I$ scattering
polarization signal, we also detected  weak $U/I$ signals, with amplitudes just
above the instrumental noise level.
 
In the next sections we describe the observations and
the reduction, and we investigate the possible presence of correlations
between the spatial variations of the polarization signals and  those
of the continuum intensity,
which we have used as an indicator of the location of granules and
intergranular lanes.

%--------------------------------------------------------------------
\section{Observations}
\label{observ}

Observations were performed on October 12$^{\rm }$, 2016, 
at the GREGOR telescope in Tenerife, using the ZIMPOL polarimeter \citep[see]
[for technical details]{ramelli-2014} , which is installed on the same
spectrograph that is used for the GREGOR Infrared Spectrograph, GRIS,
\citep{collados-2102}. The double ferroelectric crystal (FLC)
modulator \citep{gisler_2003} was installed at the entrance of the
spectrograph. 
The 1 kHz modulation allowed  us 
to simultaneously measure the full Stokes vector without
significant influences from the seeing; the polarimetric precision is thus
principally determined by the photon noise.
For the observations we took advantage of the recently installed image
derotator, which allowed us to perform long-exposure observations. 
The observations were performed close to the east limb, at $\mu = \cos
\theta \simeq 0.3$ (where $\theta$ is the heliocentric angle).
The solar surface was thus observed under an angle of $\simeq 70~\deg$.

Small pores were present in the field of view and could be used by the Shack
Hartmann wavefront sensor of the adaptive optics system \citep{berkefeld-2016}. 
The spectrograph slit was placed away from the pores, in a magnetically quiet
region, parallel to the closest limb. 
The seeing quality varied during the observation. 
The Fried parameter oscillated between $r_{0}=10$~cm and $r_{0} = 20$~cm. 
The solar area covered by the spectrograph slit corresponded to 0.3\arcsec\ 
(width of the slit) times 47\arcsec\ (length portion of the slit covered by 
the CCD sensor). 
Along the spatial direction, a ZIMPOL image has
140 pixels, so that one pixel row covered $0.33\arcsec$.
Taking into account stray light in the telescope, residual effects that
could not be fully corrected by the adaptive optics, and the sampling theorem,
we estimate to have achieved a spatial resolution of about
$0.6\arcsec$.  

The GREGOR polarimetric calibration unit, GPU, 
 \citep{hofmann-2012}
was used. This device is located at the second focal point (F2) before
any folding reflection that would produce significant instrumental
polarization. 

Four single acquisitions with 1~s exposure time each are needed to 
collect the required minimum information to calculate a Stokes image with 
reduced systematic instrumental noise \citep{gisler}.
 Each Stokes image thus required 6.7~s 
(total exposure time plus the time required for 
 digitization and data transfer). 

The complete observing procedure in a selected solar region consisted of the
following steps:

\begin{itemize}
     \item        Calculation of the  ZIMPOL camera timing parameters needed
                  to reduce systematic effects 
                 that might be introduced by large instrumental polarization.
                  To this aim, the specific procedure described in 
                 \citet{ramelli-2014} was applied.
                 Measurements were made in quiet-Sun regions. 

     \item The polarimetric calibration was performed using the GPU
       located at F2; data were collected and used to determine
       the demodulation matrix. In addition, a dark current image was stored. 

     \item A flatfield observation was taken by moving the telescope in an area
       away from the solar limb and avoiding active regions as
much as possible. 

     \item The scientific observation of the selected region was performed.

\end{itemize}

In order to follow the evolution of the demodulation matrix and of the
intensity flatfield image with the rotation of the telescope, all these
operations typically had to be repeated every 20 to 40 minutes.
 
A total of five series of observations were performed at $\mu \simeq 0.3 $,
but with the slit located at slightly different positions. The number 
of images stored in each series was 25, 250, 125, 125, and 125.

\section{Data reduction}
The Stokes images acquired with the ZIMPOL camera were corrected through 
the calibration matrix, which was calculated from the data collected during 
the polarization calibration procedure.
The intensity images were corrected for flatfield. 

The root mean square (rms) obtained in an area located in the continuum in the
fractional polarization images (obtained in 6.7~s) is equal to $\simeq
0.0028$, and it is mainly given by the statistical noise.
This value is comparable to the amplitude of the signals we are
interested in measuring. 
We therefore needed to average various images to improve 
the signal-to-noise ratio to preserve the spatial and spectral resolutions.

\begin{figure*}
\centering
\includegraphics[angle=90,width=14cm]{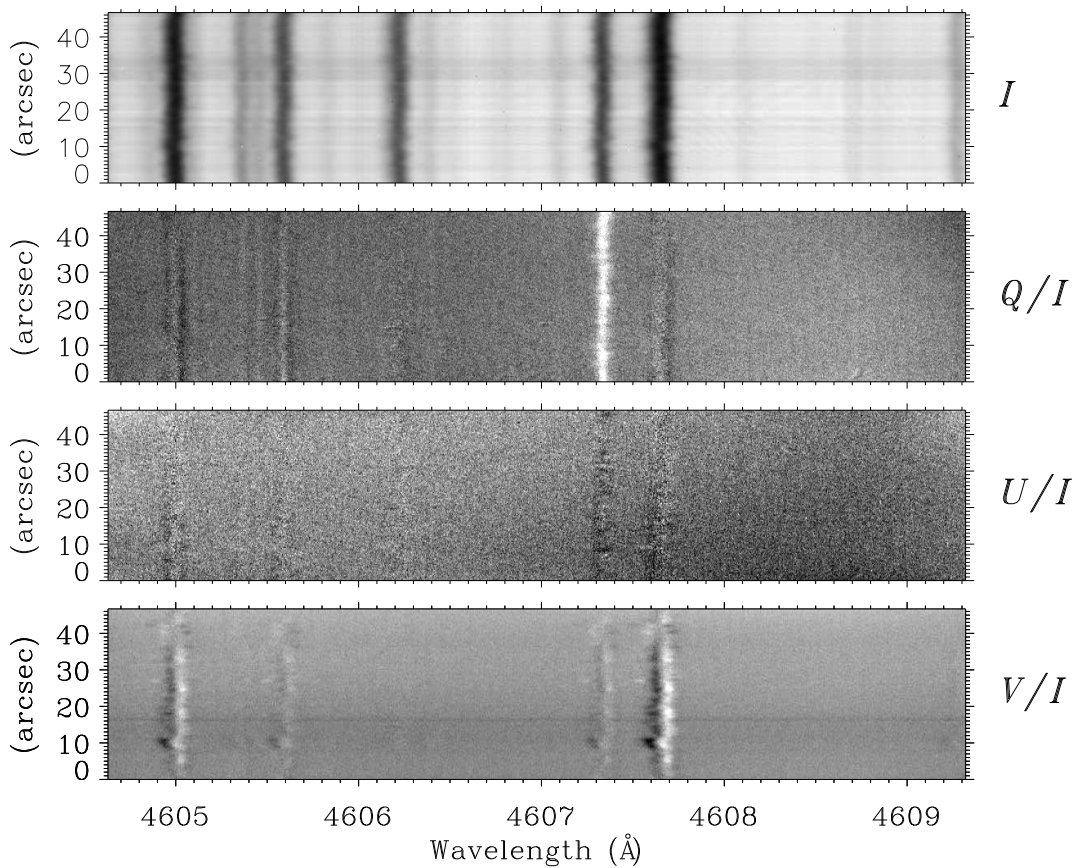}
\caption{Stokes images of a spectral interval around the Sr~{\sc i} 4607~\AA\ 
line. The spatial direction spans 47\arcsec\ on the solar disk. 
The observed region was near the est limb 
at $\mu \simeq 0.3$, and the slit was placed parallel to the
nearest limb.
 The reference direction for positive Stokes $Q$ is the tangent to the
nearest solar limb.
These images are the result of a 2-minute observation average.
The granulation pattern can be recognized in the intensity image, in
particular in the continuum. 
The  $Q/I$ image shows the scattering polarization peak in the core of the 
Sr~{\sc i} line. Spatial variations at granular scales of this peak can be
observed.
Weak $U/I$ signals, just above the noise level, can be observed in the core
of the Sr~{\sc i} line.  
In $V/I$ the typical antisymmetric Zeeman patterns can be easily recognized. 
Note in particular some small sized $V/I$ structures (for instance at spatial
position 27\arcsec).}
\label{Fig1}
\end{figure*}
 
Figure~\ref{Fig1} shows the images obtained by averaging 20
 Stokes images that were sequentially registered.
In the $I$ image we can recognize intensity variations along the spatial
direction due to the granulation. 
The $Q/I$ image shows the scattering polarization peak in the Sr~{\sc i}
4607~\AA\ line. 
The polarization peak shows clear spatial variations; detection these was in
fact the main goal of our observations.  
In the $U/I$ image it is possible to note weak signals in the Sr~{\sc i}
line core. 
Finally, in the $V/I$ image, we can recognize the typical patterns of the
longitudinal Zeeman effect.
 This image allows appreciating the high spatial resolution 
($\simeq 0.6\arcsec$) of our observation.

 Averaging 20 subsequent Stokes images, which corresponds to an integration 
over about two minutes, we were able to reach a polarimetric sensitivity of 
$\simeq 7.5 \times 10^{-4}$. A discussion of the suitability of this temporal 
average is provided in the next section.
The spectral resolution of our observation is of $\simeq 10$ m\AA.

\section{Data analysis and results}
We here focus our attention on the second series of observations
that consists of 250 images, because this series has more data.
Our goal is to analyze the spatial variations in the linear polarization in
the Sr~{\sc i} 4607~\AA\ line, 
 and to obtain some insights on the physical information that can be 
extracted from them.
In a first attempt we try to correlate the amplitude of the $Q/I$ peak with the
 continuum intensity.

\begin{figure*}
\centering
\includegraphics[angle=90,width=12cm]{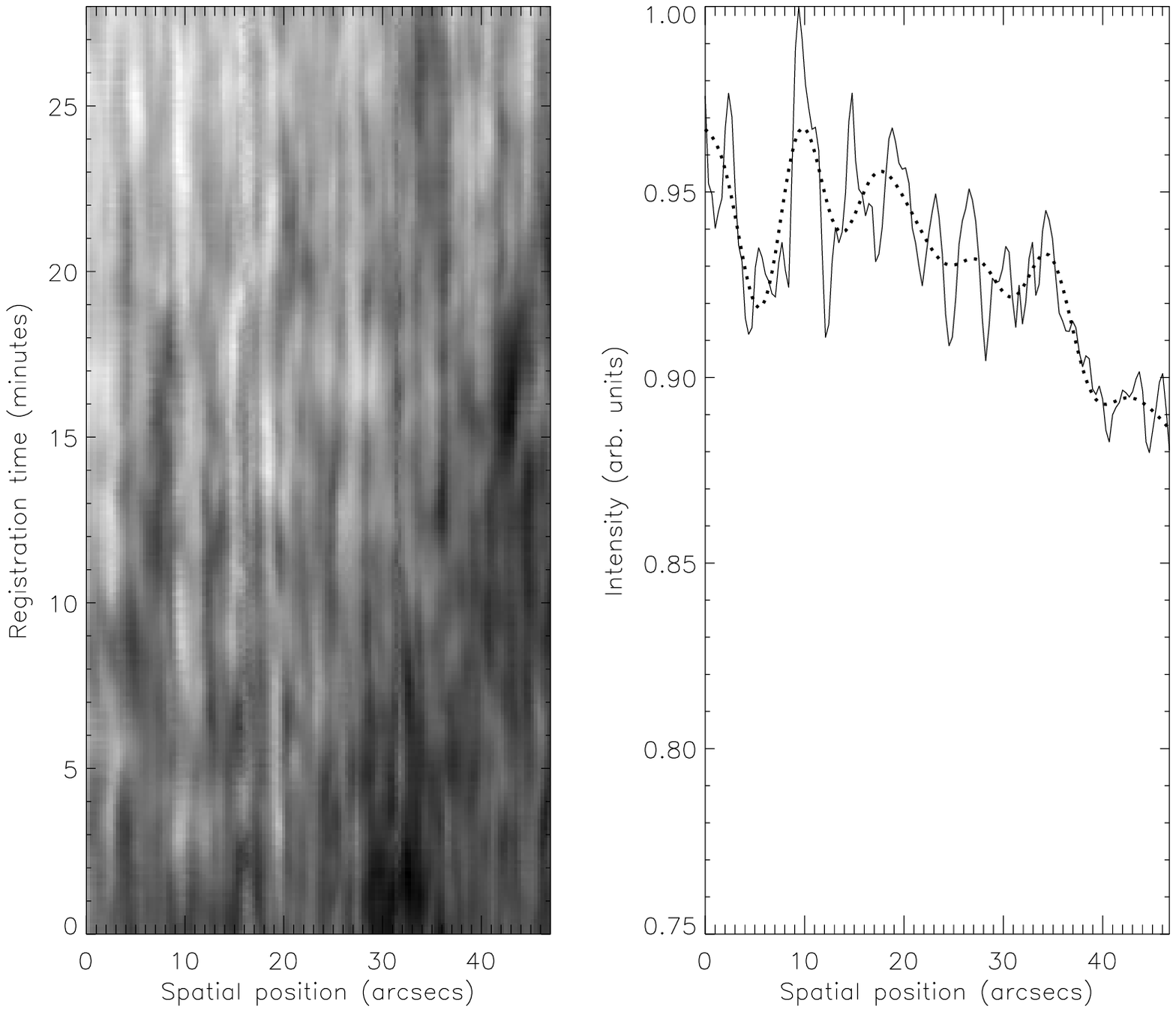}
\caption{Left panel: 
2D image composed by the 250 continuum intensity profiles measured along the
spatial direction ($47\arcsec $) and sequentially plotted along the ordinate.
The time required to store a Stokes image is 6.7~s.
 The 250 profiles thus
report the evolution of the intensity seen by the spectrograph slit in 28
minutes.
Right panel: 
one  of the 250 continuum intensity profiles (solid line).
The dotted line is a smoothed profile obtained by applying a Fourier
filter to remove fluctuations at granular scale.}
\label{Fig2}
\end{figure*}

As previously pointed out, in order to obtain an adequate signal-to-noise 
ratio in the polarization images, we averaged 20 subsequent images of the 
time series, corresponding to an integration over about two~minutes.
To evaluate whether this temporal averaging introduced significant spatial
degradation in our data, we visualized the temporal evolution of the intensity
seen by the spectrograph slit (the portion seen by the CCD sensor). 
This is shown in the left panel of Figure~\ref{Fig2}, where the continuum 
intensity is reported in gray scale as a function of the spatial position along
the slit (abscissa) and of the time (ordinate).
A single continuum intensity profile can be seen as an example in the right
panel of  the same figure (solid line).
The overplotted dotted line is the Fourier-filtered profile that reproduces 
the large-scale fluctuations,  but ``cuts'' small-scale fluctuations.

\begin{figure*}
\centering
\includegraphics[width=10cm]{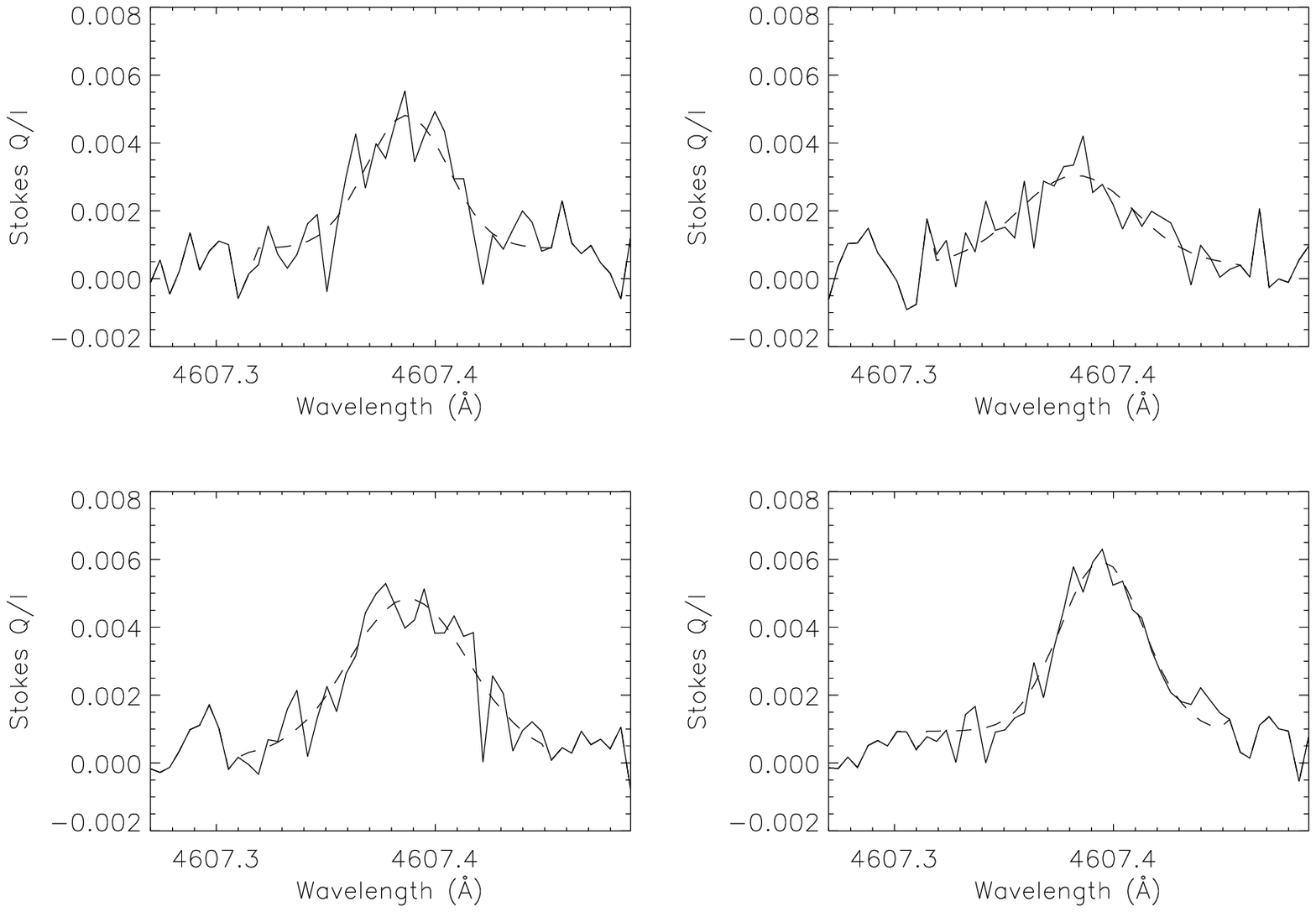}
\caption{$Q/I$ profiles in a wavelength interval around the Sr~{\sc i} line.
 The profiles correspond to a single row of pixels in the spatial 
direction.
The dashed lines are Gaussian fits of the profiles, and are used to calculate 
the polarization peak amplitude.} 
\label{Fig3}
\end{figure*}

Figure~\ref{Fig2} shows that the observed structures do not change
drastically within several minutes (as might be expected considering 
the average lifetime of the granules), thus confirming that an average 
over 20 images is suitable for our investigation.
Intensity and polarimetric data were now calculated from images obtained 
by averaging the time series of 250 Stokes images with a smoothing window width of 20 images, shifted in increments of 20 images (to avoid data 
overlap). Discarding the last 10 images of the time series, we then obtain a 
sequence of 12 averaged images.

The amplitude (as well as the width and the position) of each $Q/I$ peak in
the Sr~{\sc i} was calculated using a Gauss fit. 
Some examples of single $Q/I$ peak profiles are provided in Figure~\ref{Fig3},
showing measured profiles (solid lines) with different amplitudes, and the
corresponding Gaussian fit (dashed line).
An estimate of the errors for each fit was also calculated and stored by the
fitting procedure. 
We verified that the averaged amplitude of the $Q/I$ peaks is
compatible with the values reported by \citet{stenflobianda1997}. 

The question arises whether these variations have a solar origin.
A solar origin is supported by the following arguments. 
All the known instrumental sources of spurious signals (that might be
introduced by an imprecise optical setup) were carefully investigated and
excluded.
Moreover, the observed variations are generally larger
than the noise affecting each profile (note that the profiles in 
Figure~\ref{Fig3} are not smoothed). 
Finally, we verified that these variations disappeared when the spatial 
resolution was lower (e.g., with bad seeing conditions). 

The  continuum intensity indicates whether we observe 
a granulum or an intergranulum. 
We introduce  the following parameter: 

$\Delta I{_c} = (I{_c} -I{_{cF}}) / I{_{max,}}$

where $I{_c}$ is the continuum intensity (e.g., see the solid line in
the right panel of Figure~\ref{Fig2}), $I{_{cF}}$ is obtained with Fourier
smoothing of $I{_c}$ (dotted line in the right panel of Figure~\ref{Fig2})
that follows the large-scale fluctuation, and $I{_{max}}$ is the maximum
value in the continuum.
The parameter $\Delta I{_c}$ provides an approximate indication on
whether we observe a granulum (positive sign) or an intergranular lane
(negative sign). 

\begin{figure*}
\centering
\includegraphics[width=10cm]{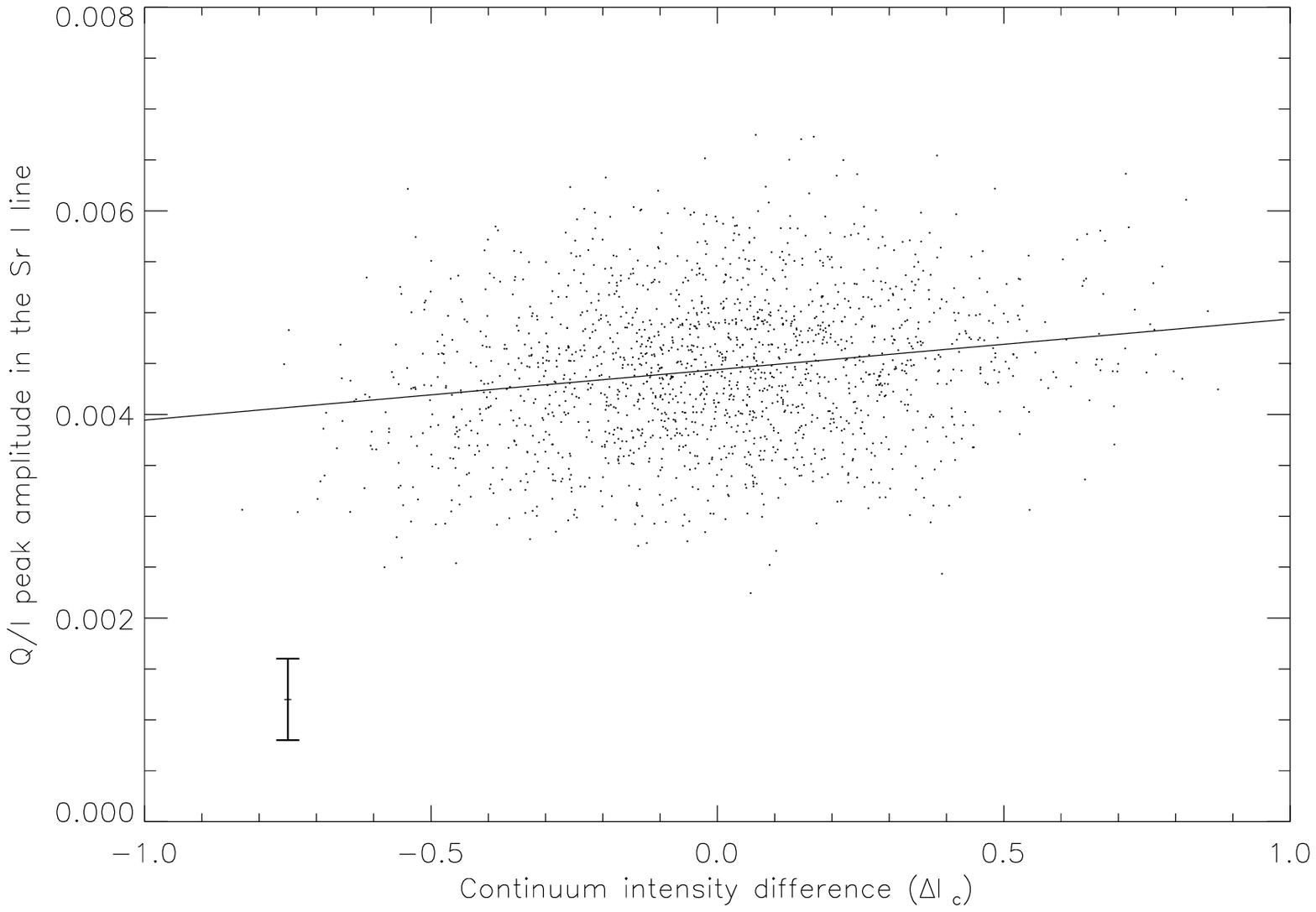}
\caption{Scatter plot of the Sr~{\sc i} $Q/I$ peak amplitude vs. the
corrected continuum intensity (see text). The error bar
 associated with each $Q/I$ value is shown
in the bottom left corner.
The solid line is a linear regression indicating a trend with a Pearson
correlation coefficient of $0.19$.} 
\label{Fig4}
\end{figure*}

The scatter plot in Figure~\ref{Fig4} shows how the $Q/I$ peak values
correlate with the $\Delta I{_c}$  parameter.  
The error in the $Q/I$ peak amplitudes is estimated by the
 Gauss fit procedure applied to the profile ($ \pm 0.0004$). 
The error bar is reported in the lower left part of Figure~\ref{Fig4}.
 
The points in the scatter plot spread over a region that is larger than the error
interval. 
A linear regression (see solid line) suggests a trend of increasing linear
polarization in the granules compared to the intergranules. The slope of 
the solid line is  
$(4.44 \pm 0.5) \times 10^{-4}$.
The Pearson correlation coefficient is $0.19$.
The trend is an increase in $Q/I$ peak amplitude toward the interior of 
the granulation cells.
 This result was also found by \citet{malherbe-2007}.

The $U/I$ panel in Figure~\ref{Fig1} shows weak signals in the
core of the Sr~{\sc i} line. 
Although their amplitudes are very small, they are expected to be of solar
origin. 
These signals can be either due to the rotation of the scattering polarization 
plane produced by the Hanle effect 
caused by
weak resolved magnetic fields or to  anisotropy
variations in the radiation field that are due to horizontal inhomogeneities 
in the photospheric plasma. 
If the Zeeman effect were to play a role, we would expect larger signatures
in the neighboring Fe~{\sc i} line, which are not observed. 
The signal-to-noise ratio in $U/I$ does not allow 
significant correlation studies between $U/I$ and other quantities, such as 
$Q/I$ or the continuum intensity.

The precision and resolution we were able to achieve in the observations
allowed us to perceive a small  
correlation between the scattering polarization amplitude and the continuum 
intensity.  
Observations at different solar positions to observe the atmosphere under
different angles would improve the information we can acquire from this
technique.
With current observing facilities, even with 
ideal observing conditions, it is not expected to achieve results with a much
better significance. 
The main limit is imposed by the noise given by the available photon 
statistics. 
Significant improvements are expected with the future 4-meter class
telescopes such as the Daniel K. Inouye Solar Telescope, DKIST, and the
European Solar Telescope, EST.

\section{Conclusions}
%\begin{enumerate}
%       \item 
Our observations allowed the detection of  spatial variations of the $Q/I$
scattering polarization  
                signal of the Sr~{\sc i} 4607~\AA\ line, measured at 
                $\mu \simeq 0.3$ (the direction of positive Stokes $Q$ is
                parallel to the nearest solar limb).  
                The spatial scale of the variations is comparable to the 
                granular scale, thus their origin has to be searched for in
                physical 
                effects that occur at this level. 
                We also detected small spatial scale $U/I$ signatures, but with
                polarimetric amplitudes that are just above the noise level
                and cannot 
                be easily analyzed. 

These small-scale signatures might be produced  by the 
            Hanle effect (depolarization and the rotation of the scattering 
            polarization plane due to oriented magnetic fields) by
            geometrical or dynamic anisotropy fluctuations of the radiation
            field. 

We find a small correlation between the scattering polarization peak 
           amplitudes of the Sr~{\sc i} 4607~\AA\ line and the continuum
           intensity.  This suggests that the polarization inside
               granulation cells is higher than in the lanes. 
 
The significance of the results presented here is mainly limited by 
           the available photon statistics. Significant improvements in this
           sense 
                are expected from the next generation of solar telescopes such
                as the DKIST and the EST.   
%\end{enumerate}

\begin{acknowledgements}
The 1.5-meter GREGOR solar telescope was built  by  a  German consortium
under  the  leadership  of  the Kiepenheuer-Institut f\"ur Sonnenphysik in
Freiburg with the Leibniz-Institut f\"ur Astrophysik Potsdam, the Institut
f\"ur Astrophysik G\"ottingen, and the Max-Planck-Institut f\"ur
Sonnensystemforschung in G\"ottingen as partners, and with contributions by
the Instituto de Astrofsica de Canarias and the Astronomical Institute of the
Academy of Sciences of the Czech Republic. 
IRSOL is supported by the Swiss Confederation (SEFRI), Canton Ticino, the
city of Locarno and the local municipalities. 
This research work was financed by SNF grants 200020\_157103  and  
200020\_169418. 
\end{acknowledgements}

\bibliographystyle{aa} % style aa.bst
\bibliography{PaperSrtoADS.bib} % your references Yourfile.bib

\end{document}